\documentclass{amsart}
 \newtheorem{thm}{Theorem}[section]
 \newtheorem{cor}[thm]{Corollary}
 \newtheorem{lem}[thm]{Lemma}
 \newtheorem{prop}[thm]{Proposition}
 \theoremstyle{definition}
 
 \theoremstyle{remark}
 
 \newtheorem*{ex}{Example}
 \numberwithin{equation}{section}

\usepackage{amssymb,amsmath}
\usepackage{amsfonts}

\newcommand{\rank}{{\rm rank}}
\newcommand{\Ker}{{\rm ker}}

\newcommand{\oR}{\mathbb R}

\newcommand{\oN}{\mathbb N}

\newcommand{\oK}{\mathbb K}
\newcommand{\DD}{\mathcal D}
\newcommand{\dd}{\mathbf d}
\newcommand{\tL}{{\tilde \Lambda}}

 \newcommand{\xx}{\mathbf x}

\newcommand{\Span}[1]{{\rm Span}(#1)}
\newcommand{\MM}{\mathcal M}
\newcommand{\PR}{\oR[\xx]}
\newcommand{\PK}{\oK[\xx]}
\newcommand{\CC}{{\mathcal C}}

\newcommand{\BB}{{\mathcal B}}

\newcommand{\AAA}{{\mathcal A}}
\newcommand{\br}{{, }}

\begin{document}

%
%
%
%
%
%
%
%
%
\title[A Sparse Flat Extension Theorem  for Moment Matrices]
 {A Sparse Flat Extension Theorem for Moment Matrices}

\author[M. Laurent]{Monique Laurent}
\address{Monique Laurent, Centrum  Wiskunde \& Informatica (CWI)\br
Kruislaan 413\br
1098 SJ Amsterdam\br
The Netherlands}
\email{M.Laurent@cwi.nl}

\author[B.~Mourrain]{Bernard Mourrain}
\address{Bernard Mourrain, GALAAD, INRIA M\'editerran\'ee\br
 BP 93, 06902 Sophia Antipolis\br
 France}
\email{mourrain@sophia.inria.fr}


\subjclass{Primary 30E05; Secondary 12D10}
\keywords{Truncated moment problem, moment matrix, Hankel operator, polynomial optimization}

\date{December 12, 2008}

\begin{abstract}
In this note we  prove a generalization of the flat extension theorem of Curto and Fialkow \cite{CF96} for truncated moment matrices. It applies to moment matrices indexed by an arbitrary set of monomials and its border,
assuming that this set is connected to 1.
 When formulated in a basis-free setting, this
 gives an equivalent result for truncated Hankel operators.
\end{abstract}

\maketitle


\section{Introduction}

Throughout this note, $\oK$ denotes a field, 
$\oK[\xx]=\oK[x_1,\ldots,x_n]$ is the ring of multivariate polynomials in $n$
variables $\xx=(x_1,\ldots,x_n)$ with coefficients in $\oK$,
$\MM_n=\{\xx^\alpha:=x_1^{\alpha_1}\cdots
x_n^{\alpha_n}\mid \alpha\in\oN^n\}$ is the set of monomials in
the variables $\xx$, 
and $\MM_{n,t}$ (resp., $\PK_t$) is the set of monomials  (resp., of polynomials) 
of degree at most $t$. The dual basis of $\MM_{n}$ in the dual space
$\PK^*$ is denoted as
$\DD_n=\{\dd^\beta\mid \beta\in \oN^n\}$.
The natural action of $\PK$ on $\PK^*$ is denoted by 
$$ (p,\Lambda)\in \PK\times \PK^* \mapsto p\cdot \Lambda\in\PK^*$$
where $(p\cdot\Lambda)(q):= \Lambda(pq)$ for $q\in \PK$.

\subsection{The moment problem} In this section, we consider $\oK=\oR$.
The moment problem (see e.g. \cite{Ak,Fu}) deals with the characterization of
the sequences of moments of measures. Given a probability measure $\mu$ on
$\oR^n$, its moment of order $a=\xx^\alpha \in\MM_n$ is the quantity $\int
\xx^\alpha\mu(dx)$. The moment problem concerns the characterization of the
sequences $y=(y_a)_{a\in\MM_n} $ that are the sequences of moments of some
nonnegative measure $\mu$, in which case one says that $\mu$ is a
representing measure for $y$, with $y_1=1$ if $\mu$ is a probability measure.
Let $\Lambda\in \PR^*$ denote the linear form on $\PR$ associated to the
sequence $y$, defined by $\Lambda(p) =\sum_a p_a y_a$ for any polynomial
$p=\sum_{a\in\MM_n}p_a a\in\PR$. Then, $y$ has a representing measure $\mu$
precisely when $\Lambda$ is given by $\Lambda(p)=\int p(x)\mu(dx)$ for all
$p\in \PR$. A well known necessary condition for the existence of a
representing measure is the positivity of $\Lambda$, i.e.  $\Lambda(p^2)\ge
0$ for all $p\in \PR$, which is equivalent to requiring that the matrix
$M(y):=(y_{ab})_{a,b\in\MM_n}$ be positive semidefinite.  As is well known
this necessary condition is also sufficient in the univariate case ($n=1$)
(Hamburger's theorem), but it is not sufficient in the multivariate case ($n\ge
2$).  However, positivity is sufficient for the existence of a representing
measure under some additional assumptions. This is the case, for instance,
when the sequence $y$ is bounded \cite{BCR} or, more generally, exponentially
bounded \cite{BM}.  The next result of Curto and Fialkow \cite{CF96} shows
that this is also the case when the matrix $M(y)$ has finite rank (cf. also
\cite{Lau05,Lau08} for a short proof).

\begin{thm}\label{theorank} \cite{CF96}
If  $M(y)$ is positive semidefinite and 
the rank of $M(y)$ is finite, then $y$ has a (unique) representing measure (which is finitely atomic with $\rank\ M(y)$ atoms).
\end{thm}

In the univariate case $n=1$, a matrix of the form $M(y)$ is a Hankel matrix. In the multivariate case, $M(y)$ is known as a 
{\em     generalized Hankel  matrix} (see \cite{MouPan})
or {\em moment matrix} (see \cite{Lau08}).
One can also define {\em truncated} moment matrices:
A matrix $M$ indexed by a subset $\CC\subseteq \MM_n$
is said to be a {\em moment matrix} if $M_{a,b}=M_{a',b'}$ for all $a,b,a',b'\in \CC$ with $ab=a'b'$. Thus its entries are given by a sequence 
$y=(y_c)_{c\in \CC\cdot \CC}$, where
$\CC\cdot\CC:=\{ab\mid a,b\in\CC\}$, and we can write $M=M_\CC(y)$. 
When $\CC=\MM_{n,t}$, we also write $M=M_t(y)$, where the entries of $y$ are indexed by
$\MM_{n,2t}$.
Such matrices 
arise naturally in the context of  the truncated moment problem, 
which asks  for the existence of a representing measure for a truncated sequence indexed by a subset 
of monomials. 
A solution to the truncated moment problem would in fact imply a solution to the moment problem. Indeed, Stochel \cite{Sto}
 shows that a sequence 
$y=(y_a)_{a\in\MM_n}$ has a representing measure if and only if 
the truncated sequence $(y_a)_{a\in\MM_{n,t}}$ has a representing measure for all $t\in \oN$.

\subsection{The flat extension theorem of Curto and Fialkow}
Curto and Fialkow studied intensively the truncated moment problem (cf. e.g. \cite{CF96,CF98,CF00} and further references therein).
In particular, they observed that the notion of {\em flat extension} of matrices 
plays a central role in this problem.
Given matrices $M_\CC$ and $M_\BB$ indexed, respectively,
 by $\CC$ and $\BB\subseteq \CC$,
$M_\CC$ is said to be a {\em flat extension} of $M_\BB$ if $M_\BB$ coincides with the principal submatrix
of $M_\CC$ indexed by $\BB$ and $\rank \ M_\CC=\rank\ M_\BB$.
Curto and Fialkow \cite{CF96} show the following 
result for truncated moment matrices.

\begin{thm}[The flat extension theorem \cite{CF96}]
 \label{theoflatCF}
For a  sequence $y=(y_a)_{a\in\MM_{n,2t}}$, if
$M_t(y)$ is a flat extension of $M_{t-1}(y)$, then there exists a (unique)
sequence $\tilde y=(\tilde y_a)_{a\in\MM_n}$ for which $M(\tilde y)$
is a flat extension of $M_t(y)$.
\end{thm}
The flat extension theorem combined with Theorem \ref{theorank} directly implies
the following sufficient condition for existence of a representing measure.

\begin{cor}\label{corCF}
 For a sequence $y=(y_a)_{a\in \MM_{n,2t}}$, if
 $M_t(y)$ is positive semidefinite and $M_t(y)$ is a flat extension of $M_{t-1}(y)$,
 then $y$ has a representing measure.
\end{cor}

 Curto and Fialkow \cite{CF98} show moreover that the flat extension condition is in some sense necessary and sufficient for 
the existence of a representing measure. More precisely, they show that
  a sequence $y=(y_a)_{a\in\MM_{n,2t}}$
has a representing measure if and only if it can be extended to 
a sequence $y'=(y'_a)_{a\in\MM_{n,2t+2k+2}}$ (for some $k\ge 0$) for which
$M_{t+k+1}(y')$ is a flat extension of $M_{t+k}(y')$.

\medskip
The proof of Theorem \ref{theoflatCF} relies on a ``truncated ideal like" property
of the kernel of flat moment matrices (see \eqref{obs1} below).
This permits to set up a  linear system of equations 
in order to construct the flat extension 
$M_{t+1}(\tilde y)$ of $M_t(y)$ (and then iteratively the infinite flat extension 
$M(\tilde y)$). 
See also \cite{Lau08} for an exposition of this proof.
Schweighofer \cite{Schw} proposes an alternative proof which is less technical and relies on properties of Gr\"obner bases.
 We propose in this note another simple alternative proof, which applies more generally to 
truncated moment matrices indexed by (suitable) general monomial sets (see Theorem \ref{mytheoflat}). 

\subsection{A generalized flat extension theorem}

We need some definitions to state our extension of Theorem \ref{theoflatCF}.
For  $\CC\subseteq \MM_n$, 
$$\CC^+:=\CC\cup\ \bigcup_{i=1}^n x_i\CC=\{m,x_1m,\ldots,x_nm\mid m\in \CC\} 
\ \text{ and } \
\partial \CC:=\CC^+\setminus \CC$$
are called, respectively, the 
{\em closure} and the {\em border} of  $\CC$.
The set $\CC\subseteq \MM_n$ is said to be {\em connected to 1} 
if $1\in\CC$ and every monomial $m\in\CC\setminus \{1\}$ can be written as $m=x_{i_1}\cdots x_{i_k}$
with $x_{i_1},x_{i_1}x_{i_2}, \ldots, x_{i_1}\cdots x_{i_k}\in \CC$.
For instance, $\CC$ is connected to 1 if $\CC$ is closed under taking divisions. 
For example,
 $\{1,x_2,x_1x_2\}$ is connected to 1 but  $\{1,x_1x_2\}$ is not.
We now state our main result.

\begin{thm}\label{mytheoflat}
Consider a sequence $y=(y_a)_{a\in \CC^+\cdot \CC^+}$, where $\CC\subseteq \MM_n$
is finite and connected to 1. 
If $M_{\CC^+}(y)$ is a flat extension of $M_\CC(y)$,
then there exists a (unique) sequence $\tilde y=(\tilde y)_{a\in\MM_n}$ 
for which $M(\tilde y)$ is a flat extension of $M_{\CC^+}(y)$.
\end{thm}

The proof is delayed till Section \ref{secproof}. 
Note that Theorem \ref{theoflatCF} follows directly from
 Theorem \ref{mytheoflat} applied to the case $\CC=\MM_{n,t-1}$.
Thus our result can be seen as a {\em sparse} version of Theorem~\ref{theoflatCF},
which applies to a more general monomial set $\CC$, not necessarily the full set of monomials up to a given degree.
We now give an example showing that the assumption that $\CC$ is connected to 1 cannot be omitted.

\begin{ex} 
For $n=1$, consider the set $\CC=\{1,x^3\}$, which is not connected to 1, with $\partial
\CC=\{x,x^4\}$. Consider the sequence $y\in \oR^{\CC^+\cdot\CC^+}$ defined by
$y_1=y_{x}=y_{x^2}=1$, $y_{x^3}=y_{x^4}=y_{x^5}=a$ and
$y_{x^6}=y_{x^7}=y_{x^8}=b$, where $a,b$ are scalars with $b\ne a^2$.
Then, $\rank \ M_{\CC^+}(y)=\rank\ M_{\CC}(y)=2$. If 
there is a flat extension $M(\tilde y)$ of $M_{\CC^+}(y)$, then its
 principal submatrix indexed by 
$\CC^+\cup\{x^2\}$ has the form:
$$
M_{\CC^+\cup\{x^2\}}(\tilde y)= \bordermatrix{ & 1&  x^3 & x & x^4 & x^2 \cr
1 &    1 & a & 1 & a & 1\cr
x^3 &  a & b & a & b & a \cr
x &   1 & a & 1 & a & a\cr
x^4 &   a & b & a & b & b\cr
x^2 &   1 & a & a & b & a
}$$
However, $1-x\in \ker M_{\CC^+}(y)$ implies $x-x^2\in \ker M_{\CC^+\cup\{x^2\}}(\tilde y)$ (see \eqref{obs1}) and
thus $1=a=b$, contradicting  our choice $b\ne a^2$.
Hence no flat extension exists.
\end{ex}

\subsection{Basis-free reformulation}
Here we reformulate our result in a basis-free setting. Moment matrices correspond indeed to choosing the monomial basis
$\MM_n$ in the polynomial ring $\PK$ and its dual basis $\DD_n$ in the dual space
$\PK^*$. 
Given $\Lambda\in\PK^*$, the operator 
$$\begin{array}{llll}
H_\Lambda: & \PK & \rightarrow & \PK^* \\
           & p & \mapsto & p\cdot \Lambda
\end{array}$$
is known as a {\em Hankel operator}. Its matrix with respect to the bases $\MM_n$ and $\DD_n$ is precisely the moment matrix
$(\Lambda(x^{\alpha+\beta}))_{\alpha,\beta\in\MM_n}=M(y)$ of the sequence
$y=(\Lambda(a))_{a\in\MM_n}$. 
The  kernel of $H_\Lambda$,
$$\Ker \ H_\Lambda=\{p\in \PK\mid \Lambda(pq)=0\ \forall q\in \PK\},$$
is an ideal in $\PK$.
Moreover, when $\oK=\oR$ and $\Lambda$ is positive, i.e.
when $\Lambda(p^2)\ge 0$ for all $p\in \PR$, $\Ker \ H_\Lambda$ is a real
radical ideal \cite{Lau05}.
Theorem \ref{theorank} means that $\Lambda\in\PR^*$ is positive with $\rank \ H_\Lambda <\infty$
if and only if there exists a nonnegative finite atomic  measure $\mu$ for which 
$\Lambda(p)=\int p(x)\mu(dx)$ for all $p\in \PR$.

Truncated Hankel operators can be analogously defined.
Given $\CC\subseteq \MM_n$ and
$\Lambda\in (\Span{\CC^+\cdot \CC^+})^*$, 
the corresponding Hankel operator is
$$\begin{array}{llll}
H_\Lambda^{\CC^+}: & \Span{\CC^+} & \rightarrow & \Span{\CC^+}^* \\
                   & p & \mapsto & p\cdot \Lambda 
\end{array}
$$
and its restriction to $\Span{\CC}$ is 
$H_\Lambda^\CC:\Span{\CC}\rightarrow \Span{\CC}^*$.
We have the following mappings:
\begin{equation}\label{map}
\frac{\Span{\CC}}{\Ker \ H_\Lambda^\CC}\  \stackrel{\sigma_1}{\longleftarrow} \ 
\frac{\Span{\CC}}{\Ker \ H_\Lambda^{\CC^+}\cap \Span{\CC}} \ 
\stackrel{\sigma_2}{\longrightarrow}\
\frac{\Span{\CC^+}}{\Ker \ H_\Lambda^{\CC^+}}
\end{equation}
where $\sigma_1$ is onto and $\sigma_2$ is one-to-one, so that
\begin{equation}\label{dim}
\dim\ \frac{\Span{\CC}}{\Ker \ H_\Lambda^\CC}
\ \le \  \dim \ \frac{\Span{\CC}}{\Ker \ H_\Lambda^{\CC^+}\cap \Span{\CC}}
\ \le\  \dim \ \frac{\Span{\CC^+}}{\Ker \ H_\Lambda^{\CC^+}}.
\end{equation}
Thus, $\rank \ H_\Lambda^{\CC^+}=\rank\ H_\Lambda^{\CC}$ (in which case we also say that
$ H_\Lambda^{\CC^+}$ is a  flat extension of $H_\Lambda^{\CC}$)
if and only if equality holds throughout in \eqref{dim}, i.e. both $\sigma_1$ and $\sigma_2$ in 
\eqref{map} are isomorphisms or, equivalently, if
$$\Span{\CC^+}=\Span{\CC} + \Ker H_\Lambda^{\CC^+} \ 
\text{ and } \
\ \Ker \ H_\Lambda^\CC= \Ker \ H_\Lambda^{\CC^+}  \cap \Span{\CC}.
$$
Theorem \ref{mytheoflat} can be reformulated as follows.

\begin{thm}\label{mytheoflat1}
Let $\Lambda \in (\Span{\CC^+\cdot \CC^+})^*$, where 
$\CC\subseteq \MM_n$ is finite and connected to 1, and
assume that $\rank \ H_\Lambda^{\CC^+}= \rank \ H_\Lambda^\CC$.
Then there exists (a unique) $\tL\in \PR^*$ for which $H_{\tL}$ is a flat extension of
$H_\Lambda^{\CC^+}$, i.e. $\tL$ coincides with $\Lambda$ on $\Span{\CC^+\cdot \CC^+}$ and
$\rank\ H_\tL=\rank \ H_\Lambda^{\CC^+}$.
\end{thm}

\subsection{Border bases and commuting multiplication operators}
We recall here a result of \cite{Mou99} about border bases of polynomial ideals 
that we exploit to prove our flat extension theorem.
Let $\BB:=\{b_1,\ldots,b_N\}$ be a finite set of monomials. 
Assume that, for each border monomial $x_ib_j\in \partial \BB$, we are given a polynomial
of the form
$$g^{(ij)}:= x_ib_j-\sum _{h=1}^N a^{(ij)}_h b_h\ \ \text{ where } \ a^{(ij)}_h\in\oK.$$
The set
\begin{equation}\label{setF}
F:=\{g^{(ij)}\mid i=1,\ldots,n,\ j=1,\ldots,N \ \text{ with } 
x_ib_j\in\partial \BB\}\end{equation}
is known as a {\em border prebasis} \cite{KKR} or a {\em rewriting family} for $\BB$ \cite{Mou99}.
When the set $\BB$ contains the constant monomial 1, one can easily verify
that $\BB$ is a generating set for the quotient space $\PK/(F)$, where $(F)$ is the ideal generated by the set $F$.
When $\BB$ is connected to 1, Theorem \ref{theocommute} below characterizes  the case when $\BB$ is a basis of
$\PK/(F)$, in which case  $F$ is said to be a {\em border basis} of the ideal $(F)$.
For this, for each $i=1,\ldots,n$,  consider the linear operator:
\begin{equation}\label{eqchi}
\begin{array}{llll}
\chi_i: & \Span{\BB} & \rightarrow & \Span{\BB}\\
        & b_j & \mapsto & 
\chi_i(b_j) = \left\{ 
\begin{array}{ll}
x_ib_j & \text{ if } x_ib_j\in\BB,\\
\sum_{h=1}^Na^{(ij)}_hb_h & \text{ if } x_ib_j\in \partial \BB
\end{array}\right.
\end{array}
\end{equation}
extended to $\Span{\BB}$ by linearity.
When $\BB$ is a basis of $\PK/(F)$,  $\chi_i$ corresponds to  the 
``multiplication operator by $x_i$" from $\PK/(F)$ to
$\PK/(F)$ and thus the operators $\chi_1,\ldots,\chi_n$ commute pairwise. 
The next result of \cite{Mou99} shows that the converse implication holds
when $\BB$ is connected to 1; this was also proved later in \cite{KKR} when
$\BB$ is closed under taking divisions. 

\begin{thm}\label{theocommute}\cite{Mou99}
Let $\BB\subseteq \MM_n$ be a finite set of monomials which is connected to
1, let $F$ be a rewriting family for $\BB$ as in \eqref{setF}, and let 
$\chi_1,\ldots,\chi_n$ be defined as in \eqref{eqchi}.
The set $\BB$ is a basis of the quotient space $\PK/(F)$ if and only if the
operators  $\chi_1,\ldots,\chi_n$ commute pairwise.
\end{thm}
The proof of our sparse flat extansion theorem is an adaptation of this
result to kernels of Hankel operators, where we omit 
the assumption that $B$ is connected to $1$.

\subsection{Contents of the paper}
Section \ref{secproof} contains the proof of our generalized flat extension theorem 
and we mention some applications in Section \ref{secappli}.
In particular, we observe that 
Theorem \ref{theoflatCF} is an `easy' instance of our flat extension theorem
(since one can prove existence of a basis  connected to 1). 
We also point out the relevance of the flat extension theorem to polynomial optimization
and to the problem of computing real roots to systems of polynomial equations.

\section{Proof of the flat extension theorem}\label{secproof}

We give here the proof of Theorem \ref{mytheoflat1} (equivalently, of Theorem \ref{mytheoflat}).
We will often use the following simple observations,
 which follow directly from the assumption that $\rank \ H_\Lambda^{\CC^+}=
 \rank \ H_{\Lambda}^\CC$: For all $p\in \Span{\CC^+}$, 
\begin{equation}\label{obs}
p\in \Ker \ H_\Lambda^{\CC^+} 
\stackrel{\text{def.}}{\Longleftrightarrow}
 \Lambda(ap)=0 \ \forall a\in \CC^+ 
\Longleftrightarrow
\Lambda(ap)=0 \ \forall a\in \CC,
\end{equation}
\begin{equation}\label{obs1}
p\in\Ker\ H_\Lambda^{\CC^+}\ \text{ and } \ x_ip\in\Span{\CC^+} \
\Longrightarrow \ 
x_ip\in \Ker \ H_\Lambda^{\CC^+}.
\end{equation}
Our objective  is to construct a linear form $\tL\in\PK^*$ whose Hankel operator $H_{\tL}$ is 
a flat extension of $H_\Lambda^{\CC^+}$.

Let $\BB\subseteq \CC$ for which  $\rank \ H_\Lambda^{\CC^+}=\rank \ H_\Lambda^\BB=|\BB|$. Note that we can assume that $1\in \BB$. Indeed, if no such
 $B$ exists containing 1, then 
$\Lambda(p)=0$  $\ \forall p\in \Span{\CC^+}$ and one can easily verify that this implies that $\Lambda$ is identically zero, in which case the theorem trivially holds.

\medskip
From the assumption: $\rank \ H_\Lambda^{\CC^+}=\rank \ H_\Lambda^\BB=|\BB|$, we 
have the direct sum decomposition:
$\Span{\CC^+}=\Span{\BB}\oplus \Ker \ H_\Lambda^{\CC^+}$,
and thus
\begin{equation}\label{eqm}
\forall p\in \Span{\CC^+} \ \ \  \ \exists !\ \pi(p)\in \Span{\BB} \ \text{ such that  }\
f(p):= p-\pi(p) \in \Ker \ H_\Lambda^{\CC^+}.
\end{equation}
Then the set
$$F:=\{f(m)=m-\pi(m)\mid m\in\partial \BB\}$$
is a rewriting family for $\BB$ and, for $i=1,\ldots,n$, the linear operator 
$\chi_i$ in  \eqref{eqchi} maps $p\in \Span{\BB}$ to $\chi_i(p)=\pi(x_ip)\in \Span{\BB}$.
We show that  $ \chi_1,\ldots,\chi_n$ commute pairwise.
Set $K:=\Ker\ H_\Lambda^{\CC^+}$.

\begin{lem}\label{lemcommute}
$\chi_i \circ \chi_j=\chi_j\circ \chi_i.$
\end{lem}

\begin{proof}
Let $m\in\BB$. Write $\pi(x_im):=\sum_{b\in\BB}\lambda_b^i b$ ($\lambda_b^i\in\oR$).
We have:
$$
\begin{array}{c}
\chi_j\circ\chi_i(m)
=\chi_j(\sum_{b\in\BB}\lambda_b^i b) = \sum_{b\in\BB}\lambda_b^i \chi_j(b)
=  \sum_{b\in\BB}\lambda_b^i  (x_jb-f(x_jb))\\
= x_j(\sum_{b\in\BB}\lambda_b^i b) -  \sum_{b\in\BB}\lambda_b^i f(x_jb) 
= x_j(x_im-f(x_im)) - \sum_{b\in\BB}\lambda_b^i f(x_jb).
\end{array}
$$
Therefore,
$$\begin{array}{l}
p:= \chi_j\circ\chi_i(m) - \chi_i\circ\chi_j(m)
= \underbrace{x_if(x_jm)-x_jf(x_im)}_{p_1} 
+ \underbrace{\sum_{b\in\BB}\lambda_b^jf(x_jb)-\lambda_b^if(x_ib)}_{p_2}.
\end{array}
$$
We show that  $p_1\in  K$. Indeed, 
$\ \forall a\in \CC$, 
$ \Lambda(ap_1) = \Lambda( ax_if(x_jm) - ax_j(x_im)) =0$
since $ax_i,ax_j\in\CC^+$ and $f(x_im), \ f(x_jm)\in K$;
by \eqref{obs}, this shows that 
$p_1\in K$.
As $p_2\in K$ too, this implies 
$p\in K$ and thus $p=0$, because 
$p\in\Span{\BB}$.
\end{proof}

Our objective now is to show that $\BB$ is a basis of $\PK/(F)$ and that, if
$\tilde \pi$ denotes the projection from $\PK$ onto $\Span{\BB}$ along $(F)$, then  the  operator $\tL$ defined
by $\tL(p)=\Lambda(\tilde \pi(p))$ for $p\in\PK$, defines the desired
 flat extension of $\Lambda$.
Note that when $\BB$ is connected to 1, Theorem \ref{theocommute} implies directly that $\BB$ is a basis of $\PK/(F)$. 
As we do not assume $\BB$ connected to 1, we cannot apply Theorem \ref{theocommute}, but our arguments below are inspired from its proof.
In particular, we construct the projection $\tilde \pi$ via the mapping $\varphi$ from \eqref{phi} below.
 
\medskip
As the $\chi_i$'s commute, the operator 
$f(\chi):=f(\chi_1,\ldots,\chi_n)$ is well defined for any polynomial
$f\in\PK$. Then $\PK$ acts on $\Span{\BB}$ by
$$(f,p)\in \PK\times \Span{\BB}\mapsto 
f(\chi)(p)\in \Span{\BB}.$$
Recall that $1\in  B$. The mapping
\begin{equation}\label{phi}
\begin{array}{llll}
\varphi: & \PK & \rightarrow & \Span{\BB} \\ 
         & f & \mapsto & f(\chi)(1)
\end{array}
\end{equation}
is a homomorphism and, by the following property,
\begin{equation}\label{homo}
\varphi(fg)= f(\chi)(g(\chi)(1))=f(\chi)(\varphi(g)) \ \forall f,g\in\PK,
\end{equation}
$\Ker\ \varphi$ is an ideal in $\PK$. 
We now prove that $\varphi$ coincide on $\Span{\CC^{+}}$ with the projection $\pi$ on
$\Span{\BB}$ along $K=\ker H_{\Lambda}^{\CC^{+}}$.
\begin{lem} \label{lem:phipi}
For any element $m\in \CC^{+}$, $\varphi(m)= \pi(m)$.
\end{lem}
\begin{proof}
We use induction on the degree of $m$. If $m=1$, we have
$\varphi(1)=\pi(1)=1$ since $1\in \BB$. Let $m\neq 1 \in \CC^{+}$.
As $\CC$ is connected to 1, $m$ is of the form 
$m=x_{i} m_1$ for some $m_1\in \CC^{+}$. By the induction assumption, we have
$\varphi(m_1)=\pi(m_1)$. Then,
$$ 
\varphi(m)= \varphi(x_im_1)= 
\chi_{i} (\varphi(m_1)) = \chi_{i} (\pi(m_1) )= x_{i} \pi(m_1) - \kappa,
$$
with $\kappa \in F \subseteq K$. But we also  have
$$ 
m= x_{i}\, m_1 = x_{i} (\pi(m_1)+ m_1-\pi(m_1)) = x_{i} \pi(m_1) + x_{i}\, \kappa_1
$$
where $\kappa_1=  m_1-\pi(m_1) \in K$. We deduce that
$$ 
m = \varphi(m) +\kappa + x_{i}\, \kappa_1 = \varphi(m)+ \kappa_2
$$
with $\kappa_2= \kappa+x_{i}\, \kappa_1 \in K^{+}\cap \Span{\CC^{+}}$.  
As $\kappa_1\in K$ and $x_i\kappa_1 \in \Span{\CC^+}$, 
we deduce using \eqref{obs} that $x_i\kappa_1\in K$, thus implying
$\kappa_2\in K$. 
As $\varphi(m)\in \Span{\BB}$, it coincides with the projection of $m$ on  
$\Span{\BB}$  along $K$. 
\end{proof}
This implies directly:
\begin{equation}\label{rela}
\varphi(b)=b, \ \  \varphi(x_ib)= \chi_i(b)\  \ \forall b\in \BB \ \ \forall i=1,\ldots,n,
\end{equation}
\begin{equation}\label{corres}
 \Lambda(pq)=\Lambda(p\ \varphi(q))= \Lambda(\varphi(p)\varphi(q)) \ \ \forall p,q\in\Span{\CC^+}.
\end{equation}

\begin{lem}\label{lemres}
For all $p,q\in\Span{\CC^{+}}$, $\Lambda(pq)= \Lambda(\varphi(pq))$.
\end{lem}

\begin{proof}
We first show by induction on the degree of $m\in \CC^{+}$ 
that 
\begin{equation}\label{ind}
\Lambda(mb)=\Lambda( \varphi(mb)) \ \ \forall b\in \BB.
\end{equation}
The result is obvious if $m=1$. Else, as $\CC^{+}$ is connected to 1, we can 
 write $m=x_i m_1$ where $m_1 \in\CC^{+}$. 
Using first \eqref{corres} and then \eqref{rela}, we find:
$$\Lambda(mb)=\Lambda(m_1x_ib) =\Lambda(m_1 \varphi(x_ib))=
\Lambda(m_1 \chi_i(b)).$$
 Next,
using first the  induction assumption and then \eqref{homo}, \eqref{rela}, we find:
$$\Lambda(m_1 \chi_i(b))= \Lambda(\varphi(m_1\chi_i(b)))
= \Lambda(m_1(\chi)(\chi_i(b)))
= \Lambda(m(\chi)(b))
=\Lambda(\varphi(mb)),$$
thus showing \eqref{ind}.
We can now conclude the proof of the lemma.
Let $p,q\in\Span{\CC^+}$. Then, using successively \eqref{corres}, \eqref{ind}, \eqref{homo}, \eqref{rela},  $\Lambda(pq)$ is equal to
$$
\Lambda(p\ \varphi(q)) 
= \Lambda(\varphi(p\varphi(q)))
=\Lambda(p(\chi)(\varphi(\varphi(q))))
= \Lambda( p(\chi)(\varphi(q))) 
= \Lambda(\varphi(pq)).\qedhere$$
\end{proof}

We can now conclude the proof of Theorem \ref{mytheoflat1}.
Let $\tL$ be the linear operator on $\PK$ defined by
$$\tL(p):=\Lambda(\varphi(p)) \ \ \text{ for } p\in \PK.$$
We show that $H_\tL$ is  the unique  flat extension of $H_\Lambda^{\CC^+}$.

First, $H_\tL$ is an extension of  $H_\Lambda^{\CC^+}$ since,
for all $p,q\in \Span{\CC^+}$, 
$\tL(pq)= \Lambda(\varphi(pq))= \Lambda(pq)$ (by Lemma \ref{lemres}).

Next, we have $K = \ker H_{\Lambda}^{\CC^{+}} \subseteq \ker
H_{\tL}$. Indeed, let  $\kappa \in K$. By Lemma~\ref{lem:phipi}, 
 $\varphi(\kappa)= \pi(\kappa)=0$.  Thus for any $p\in \PK$, we have
$\tL(p\, \kappa)= \Lambda( \varphi(p \, \kappa) )=
\Lambda(p(\chi)(\varphi(\kappa)))=0$, which shows that $\kappa \in \ker
H_{\tL}$.

As $F$ is a rewritting family for
$\BB$ and $\BB$ contains $1$,
 $\BB$ is a generating set of  $\PK/(F)$ and thus $\dim \PK/(F) \le |\BB|$.
Set  $\AAA_{\tilde{\Lambda}}:= \PK/\ker H_{\tL}$.
Then, as $F\subseteq K \subseteq \ker H_{\tL}$, we have
$\dim \AAA_{\tilde{\Lambda}} \le \dim \PK/(F)\le |\BB|$.
On the other hand, $\dim\AAA_{\tilde{\Lambda}} = \rank H_{\tL} \ge 
\rank H_{\tL}^{\BB}
= \rank H_{\Lambda}^{\BB}= |\BB|$.
Therefore, $\dim \AAA_{\tilde{\Lambda}}=\rank H_\tL = |\BB|$, $\ker H_\tL=(K)$,
$H_\tL$ is a flat extension of $H_\Lambda^{\CC^+}$,
and we have  the direct sum:
 $\PK=\Span{\BB} \oplus \Ker\ H_\tL$.
Moreover, $\varphi(p)$ is the projection of $p\in \PK$ on $\Span{\BB}$ along 
$\ker H_\tL$. Indeed, $\varphi(p)\in\Span{\BB}$ and $p-\varphi(p)\in \ker H_\tL$ for any $p\in\PK$ since,
for any $q\in \PK$, 
$$\begin{array}{l}
\tL(pq)=\Lambda(\varphi(pq))= \Lambda(p(\chi)(\varphi(q))),\\
\tL(p\varphi(q))= \Lambda(\varphi(p\varphi(q)))= 
\Lambda(p(\chi)(\varphi(q)))=\tL(pq).
\end{array}$$

Finally, if $\Lambda'\in \PK^*$ is another linear form whose Hankel operator
$H_{\Lambda'}$ is a flat extension of $H_\Lambda^{\CC^+}$, then 
$\ker H_\tL= (K)\subseteq \Ker\ H_{\Lambda'}$. This implies that
for all $p\in \PK$, $\Lambda'(p)=\Lambda'(\varphi(p))=\Lambda(\varphi(p)) 
= \tL(p)$.
This shows the unicity of the flat extension of $H_\Lambda^{\CC^+}$,
which concludes the proof of Theorem~\ref{mytheoflat1}.

\section{Applications}\label{secappli}

\subsection{Application to the flat extension theorem of Curto and Fialkow}
Theorem \ref{theoflatCF} is in some sense an `easy' instance of Theorem~\ref{mytheoflat}. Indeed, under its assumptions,
one can show existence of a maximum rank principal submatrix of $M_{t-1}(y)$ indexed by a monomial set $\BB$ connected to 1 which, as noted in the proof of Theorem~\ref{mytheoflat}, 
permits to apply Theorem \ref{theocommute}.

\begin{prop}\label{lembase}
Let $\Lambda\in (\Span{\CC^+\cdot\CC^+})^*$, where $\CC:=\MM_{n,t-1}$. If 
$\rank\ H_\Lambda^{\CC^+}=\rank \ H_\Lambda^{\CC}$, 
then there exists $\BB\subseteq \CC$ closed under taking divisions (and thus connected to 1)
for which
$\rank \  H_\Lambda^{\CC^+}=\rank\ H_\Lambda^\BB= |\BB|$.
\end{prop}

\begin{proof}
Let $M=(\Lambda(ab))_{a,b\in \CC^+}$ denote the matrix of $H_\Lambda^{\CC^+}$ in the canonical bases.
Consider a total degree monomial ordering $\preceq$ of $\CC$  and let
$\BB\subseteq \CC$ index a maximum linearly independent set of columns of $M$  which is constructed 
by the greedy algorithm using the ordering $\preceq$. 
One can easily verify that 
$\BB$ is closed under taking divisions (cf. \cite{LLR08b}).
\end{proof}


The following example shows that, even if $\CC$ is connected to $1$, there may
not always exist a base $\BB$ connected to 1 for $H_\Lambda^\CC$ (which
justifies our generalisation of Theorem \ref{theocommute} to kernels of Hankel operators).

\begin{ex}
For $n=2$, let $\CC=\{1,x_1,x_1x_2\}$ with 
$\partial \CC=\{x_2,x_1x_2^2,x_1^2,x_1^2x_2\}$, and let
$\Lambda\in (\Span{\CC^+\cdot\CC^+})^*$ be defined by
$\Lambda(x_1^ix_2^j)=1$ if $j=0,1$, and $\Lambda(x_1^ix_2^j)=a$ if $j=2,3,4$, except 
$\Lambda(x_1^2x_2^4)=a^2$, where $a$ is a scalar with $a\ne 1$. The associated moment matrix has the form
$$\bordermatrix{ & 1 & x_1 & x_1x_2 & x_1^2 & x_1^2x_2 & x_2 & x_1x_2^2 \cr
1 &          1 & 1 & 1 & 1 & 1 & 1 & a \cr
x_1&          1 & 1 & 1 & 1 & 1 & 1 & a \cr
x_1x_2&     1 & 1 & a & 1 & a & a & a \cr
x_1^2&       1 & 1 & 1 & 1 & 1 & 1 & a \cr
x_1^2x_2&     1 & 1 & a & 1 & a & a & a \cr
x_2&          1 & 1 & a & 1 & a & a & a \cr
x_1x_2^2 &   a & a & a & a & a & a & a^2 \cr
}$$
and
$\rank \ H_\Lambda^\CC=\rank\ H_\Lambda^{\CC^+}=2$. As $1-x_1\in \ker\ H_\Lambda^{\CC^+}$, the only sets indexing a column  base
for $H_\Lambda^\CC$ are $\BB=\{1,x_1x_2\}$ and $\{x_1,x_1x_2\}$, thus not connected to 1.
\end{ex}

Combining Theorem \ref{mytheoflat1} with Theorem \ref{theorank} we obtain the following extension of Corollary \ref{corCF}.

\begin{thm}\label{mycor}
Let $\Lambda \in (\Span{\CC^+\cdot\CC^+})^*$, where
$\CC\subseteq \MM_n$ is  finite and connected to 1.
Assume that $\Lambda$ is positive and that 
$\rank\ H_\Lambda^{\CC^+}= \rank\ H_\Lambda^\CC$.
Then the sequence $y=(\Lambda(a))_{a\in \CC^+\cdot \CC^+}$ has a representing measure.
\end{thm}

\subsection{Application to polynomial optimization}

We point out here  the relevance of the flat extension theorems 
to polynomial optimization and to the problem of computing the real roots to
polynomial equations. In this section, we take again $\oK=\oR$.

The truncated moment problem has recently attracted a lot of attention also within the optimization community, since
 it can be used to formulate 
semidefinite programming relaxations to polynomial optimization problems (see \cite{Las01a}).
Moreover the flat extension theorem of Curto and Fialkow permits to detect optimality
 of the relaxations and to extract global optimizers to the original optimization problem (see \cite{HL05}).
Here is a brief sketch; 
 see e.g. \cite{Lau08} and  references therein for details.

 Suppose we want to compute the infimum $p^*$ of a polynomial
$p$ over a semi-algebraic set $K$ defined by the polynomial inequalities
 $g_1\ge 0,\ldots,g_m\ge 0$.
 For any integer $t\ge \deg(p)/2$ and such that
 $t\ge  d_j:=\lceil\deg(g_j)/2\rceil$,
consider the program:
\begin{equation}\label{sdp}
p_t^*:=\inf\ \Lambda(p)\ \text{ s.t. } \Lambda\in (\PR_{2t})^*,\
\Lambda(1)=1,\ \Lambda\succeq 0,\ g_j\cdot \Lambda\succeq 0 \ (\forall j\le m).
\end{equation}
Here, $\Lambda\succeq 0$ means that $\Lambda$ is positive (i.e., $\Lambda(p^2)\ge 0$
for all $p\in \PR_t$) and
the localizing conditions $g_j\cdot\Lambda\succeq 0$
(i.e. $\Lambda(g_jp^2)\ge 0$ for all $p\in \PR_{t-d_j}$)
aim to  restrict the search for a representing measure suported by the set $K$ (cf. \cite{CF00,Las01a}). 
Using moment matrices, the program \eqref{sdp} can be formulated as 
 an instance of semidefinite programming for which efficient algorithms exist (see e.g. \cite{VB,handbook}).
We have: $p_t^*\le p^*$, with equality if
 $H_\Lambda^{\MM_{n,t}}$ is a 
flat extension of $H_\Lambda^{\MM_{n,t-d}}$ for an optimum solution $\Lambda$ to \eqref{sdp} ($d:=\max_jd_j$). 
In that case, the atoms of the representing measure (which exists by Corollary \ref{corCF}) are global minimizers of $p$ over the semi-algebraic set $K$  and they can be computed from $\Lambda$  \cite{HL05}. Moreover, 
they are {\em all} the global minimizers when $H_\Lambda^{\MM_{n,t}}$ 
has the maximum possible rank among all optimum solutions to the semidefinite program \eqref{sdp}.

As shown in \cite{LLR08}, 
the truncated moment problem also yields an algorithmic approach 
to the problem of computing the real roots to polynomial equations 
$g_1=0,\ldots,g_m=0$ (assuming their number is finite).
Indeed, this amounts to finding all global minimizers to a constant polynomial, say  $p=0$,
 over the real variety $K:=\{x\in \oR^n\mid g_j(x)=0\ \forall j=1,\ldots,m\}$.
Consider the semidefinite program \eqref{sdp} where the localizing conditions now read 
$g_j\cdot\Lambda=0 \ \forall j$.
 For $t$ large enough, the program \eqref{sdp} has a maximum rank solution which is a flat extension and thus, as noted above, all points of $K$ can be computed from this solution.
See  \cite{LLR08} for details.

A concern in this type of approach is the size of the matrices 
appearing in  the semidefinite program \eqref{sdp}.
In order to improve the practical applicability of this approach,
 it is crucial to derive semidefinite programs involving matrices of moderate sizes. For this one may want to consider moment matrices indexed by {\em sparse}
 sets of monomials instead of the full
degree levels $\MM_{n,t}$. This is where our new sparse flat extension theorem 
may become very useful.   It will be used, in particular, in \cite{MLLRT}.

The approach in \cite{LLR08} also permits to
find the real radical of the ideal generated by the polynomials $g_1,\ldots,g_m$.
Indeed, if $\Lambda\in (\PR)^*$ is positive, then the kernel
of its Hankel operator $H_\Lambda$ is 
a real radical ideal \cite{Lau05} and,
 under the conditions of Theorem \ref{mycor},
$\Ker \ H_\Lambda^{\CC^+}$ generates a real radical ideal.
These facts explain the relevance of moment matrices and Hankel operators 
to the problem of  finding the real radical of a polynomial ideal.
For instance, 
this permits to weaken the assumptions in Proposition 4.1 of \cite{LLR08} and to strengthen 
its  conclusions; more precisely, we do not need to assume the commutativity of the operators $\chi_i$'s (as this holds automatically, by Lemma \ref{lemcommute}) and we can claim that the returned ideal is real radical (by the above argument).

\end{document}